\documentclass[preprint]{aastex}
\usepackage{geometry}   
\usepackage{graphicx}   
\usepackage{amsmath}
\usepackage[percent]{overpic}
\usepackage{array}      
\usepackage{verbatim}   
\usepackage{bm}
\usepackage[export]{adjustbox}
\usepackage{mwe}
\usepackage{rotating}
\usepackage{enumerate}
\usepackage[table]{xcolor} 
\usepackage{color}
\usepackage{epsfig}     
\usepackage{url}        
\usepackage{subfigure}  
\usepackage{multirow} 
\usepackage{wrapfig}
\usepackage{float}

\begin{document}
\title{The Open Flux Problem}
\author{J.\ A.\ Linker$^1$, R.\ M.\ Caplan$^1$, C.\ Downs$^1$, P.\ Riley$^1$, Z.\ Mikic$^1$, R.\ Lionello$^1$, C.\ J.\ Henney$^2$, C.\ N.\ Arge$^3$, Y.\ Liu$^4$, M.\ L.\ Derosa$^5$, A.\ Yeates$^6$, and M.\ J.\ Owens$^7$}
\affil{$^1$Predictive Science Inc., 9990 Mesa Rim Road, Suite 170, San Diego, California 92121, USA}
\affil{$^2$Air Force Research Lab/Space Vehicles Directorate, 3550 Aberdeen Avenue SE, Kirtland AFB, NM, USA}
\affil{$^3$Science \& Exploration Directorate, NASA/GSFC, Greenbelt , MD 20771}
\affil{$^4$W. W. Hansen Experimental Physics Laboratory, Stanford University, Stanford, CA 94305, USA}
\affil{$^5$Lockheed Martin Solar and Astrophysics Laboratory, 3251 Hanover St. B/252, Palo Alto, CA 94304, USA}
\affil{$^6$Department of Mathematical Sciences, Durham University, Durham, DH1 3LE, UK}
\affil{$^7$Space and Atmospheric Electricity Group, Department of Meteorology, University of Reading, Earley Gate, P.O. Box 243, Reading RG6 6BB, UK}

\email{linkerj@predsci.com}

\date{September 2, 2017}


\begin{abstract}
The heliospheric magnetic field is of pivotal importance in solar and space physics.  The field is rooted in the Sun's photosphere, where it
has been observed for many years.  Global maps of the solar magnetic field based on full disk magnetograms  are commonly used as boundary conditions for coronal and solar wind models.  Two primary observational constraints on the models are (1) the open field regions in the model should approximately correspond to coronal holes observed in emission, and (2) the magnitude of the open magnetic flux in the model should match that inferred from in situ spacecraft measurements.  In this study, we calculate both MHD and PFSS solutions using fourteen different magnetic maps produced from five different types of observatory magnetograms, for the time period surrounding July, 2010.  We have found that for all of the model/map combinations, models that have coronal hole areas close to observations underestimate the interplanetary magnetic flux, or, conversely, for models to match the interplanetary flux, the modeled open field regions are larger than coronal holes observed in EUV emission.  In an alternative approach, we estimate the open magnetic flux entirely from solar observations by combining automatically detected coronal holes for Carrington rotation 2098 with observatory synoptic magnetic maps.  This approach also underestimates the interplanetary magnetic flux.  Our results imply that either typical observatory maps underestimate the Sun's magnetic flux, or a significant portion of the open magnetic flux is not rooted in regions that are obviously dark in EUV and X-ray emission.  

\end{abstract}

\section{Introduction}
\label{sec_intro}
The ``open'' magnetic field is that portion of the Sun's magnetic field that extends out into the heliosphere and becomes the interplanetary magnetic field
(IMF).  Open fields play a crucial role in heliophysics as the main driver of geomagnetic activity.  They also determine where solar energetic particles propagate and shield the solar system from galactic cosmic rays.  In the standard paradigm of coronal structure \citep[e.g.,][]{mackay_yeates2012,priest2014}, the open magnetic field originates primarily in coronal holes (CHs), regions of low intensity emission in EUV and X-rays \citep{bohlin1977,zirker1977}.  The regions that are magnetically closed trap the coronal plasma and give rise to the streamer belt that is prominent in coronagraph and eclipse images \citep[e.g.,][]{wangetal1997,linkeretal1999,pasachoffetal2009,rusinetal2010}.  While important questions remain \citep[e.g., what is the source of the slow solar wind?][]{kilpuaetal2016},  this picture accounts for many coronal and interplanetary observations.  

The IMF has been measured in situ for many years.  Ulysses measurements demonstrated that the magnitude of the radial IMF is nearly independent of heliographic latitude \citep{smith_balogh1995,smith_balogh2008}, implying that currents in the heliosphere are primarily confined to the heliospheric current sheet (HCS) and that the field is nearly potential everywhere else.  The consequence of these measurements is that the open magnetic flux of the Sun can be inferred from suitably averaged single point in situ measurements of the radial IMF \citep[e.g.,][]{owensetal2008a}.

The solar magnetic field has also been observed for over four decades, primarily in the photosphere.  Global magnetic maps are developed from full-disk magnetograms of the line-of-sight (LOS) photospheric magnetic field (inferred from the Zeeman splitting of measured spectral lines) and are available from ground and space-based observatories.  Using such maps as input, 
steady-state models have been successful in reproducing key spatial features of the large-scale corona and inner heliosphere, such as the location of CHs, the streamer belt, and the HCS.  The complexity of models can range from potential field source surface (PFSS) models to magnetohydrodynamic (MHD) models with realistic energy transport and sub-grid scale descriptions of heating and acceleration.  While the range of physical values that can be predicted depends on the details of the model, two basic properties can be predicted by all models:  
the magnitude of the open magnetic flux, and the open field regions at the solar surface.

If the basic paradigm of coronal structure is correct, then the magnitude of the open magnetic flux predicted by the combination of a coronal model and an observatory map should match that inferred from in situ spacecraft measurements.  Specifically, the open magnetic flux ($\Phi_{open}$) predicted by a coronal model can be expressed as a radial magnetic field strength at 1 AU (Astronomical Unit):
\begin{equation}
\label{eq_phi_open}
|B_{r 1 AU}| = \frac{|\Phi_{open}|}{4\pi r_{1 AU}^2} = \frac{1}{4\pi}\left(\frac{R_{ub}}{r_{1 AU}}\right)^2\int_{0}^{4\pi} |B_r(R_{ub},\theta,\phi)|d\Omega,
\end{equation}
where $r_{1 AU} = 215$ solar radii ($R_S$) and $R_{ub}$ is the upper boundary of the coronal calculation.  For PFSS models, $R_{ub}$ is the source surface radius ($R_{SS}$), and for MHD models it is the upper radial boundary.  The value of $|B_{r 1 AU}|$ should be approximately equal to the average value of $|B_r|$ measured by 1 AU spacecraft ($|B_{r IMF}|$).  In practice, this should be the average value over at least a solar rotation, as interplanetary magnetic fields fluctuate considerably, and observatory maps are built up over the rotation.  We would expect this approach to work reasonably well near solar minimum, when the recurrent patterns of coronal holes and fast solar wind streams \citep{zirker1977,luhmannetal2009,abramenkoetal2010} show that the large-scale underlying structure of the corona often varies slowly.  Accuracy could be more problematic near solar maximum, when the Sun's magnetic flux is rapidly evolving. 

Using Eq.~(\ref{eq_phi_open}), \citet{wang_sheeley1995} computed $|B_{r 1 AU}|$ for over two decades (1970-1993), using PFSS models (with $R_{SS}=2.5 R_S$).  They found that PFSS models using Wilcox Solar Observatory (WSO) magnetic maps could roughly match $|B_{r IMF}|$, but only if they used a latitude-dependent saturation factor derived for the Mount Wilson Observatory (MWO) magnetograph \citep{ulrich1992}.  The MWO factor multiplies low-latitude fields by a factor of nearly 4.5.  (The factor for the full disk is 4.5-2.5sin$^2(\rho)$, where $\rho$ is the center-to-limb angle.  \citealt{ulrichetal2009} updated this to 4.15-2.82sin$^2(\rho)$.)  This choice was controversial \citep{riley2007,rileyetal2014}.  \citet{svalgaardetal1978}, studying the performance of the WSO magnetograph, derived a constant saturation factor of 1.8 (later revised to 1.85).  \citet{wang_sheeley1995} based their argument primarily on the much better match they obtained with 1 AU measurements when using the MWO factor; when using the factor actually derived for the WSO instrument, the open flux was generally underestimated.  \citet{riley2007} took an alternative view and suggested that if the constant WSO factor was used, the missing flux could be accounted for by the contribution of CMEs.  \citet{wang_sheeley2015} revisited this topic and argued that the CME flux was insufficient to account for the open flux if the WSO derived factor is used, and that WSO maps with the original MWO correction factor and the additional flux estimated to be carried by CMEs provided the best match to interplanetary observations.

If the WSO maps corrected with the MWO factor accurately represent the solar magnetic field, we would expect that models using data from other magnetographs would independently be able to predict $|B_{r IMF}|$.  While not widely emphasized, a range of models/observatory maps are generally underestimating  $|B_{r IMF}|$ and/or  $|B_{IMF}|$.  For example, \citet{owensetal2008b} found that both the WSA-Enlil and MAS-Enlil models using National Solar Observatory (NSO) Kitt Peak data consistently underestimated $|B_{r IMF}|$ away from stream interfaces from 1995-2002.  The  models were the empirical Wang-Sheeley-Arge (WSA) model \citep{argeetal2003} coupled with the Enlil Heliospheric MHD model \citep{odstrcil2003} and the Magnetohydrodynamic Algorithm outside a Sphere (MAS) MHD model, also coupled with Enlil.   These models are elements of CORHEL \citep[Corona-Heliosphere,][]{rileyetal2012}.  \citet{stevensetal2012} found that CORHEL (utilizing MAS and several different observatory maps) systematically underestimated $|B_{r IMF}|$ measured at Ulysses.   \citet{jianetal2015} compared several model/observatory map combinations available at the Community Coordinated Modeling Center (CCMC) and found regular underestimates of $|B_{IMF}|$.  These issues have led to ad hoc correction factors being applied to input magnetic fields in order to obtain a better match.  The WSA-Enlil model is frequently run using the polarity of $B_r$ from the WSA model, and the magnitude replaced empirically \citep{mcgregoretal2011}.  \citet{linkeretal2016} showed that PFSS models computed daily using maps from the Air Force Data Assimilative Photospheric flux Transport (ADAPT)  model \citep{argeetal2010,hickmannetal2015}, generated with the assimilation of  NSO SOLIS (Synoptic Optical Long-Term Investigation of the Sun) Vector Spectromagnetograph (VSM) magnetograms, could capture the large variation in open magnetic flux seen in OMNI in situ measurements from 2003 to 2008 - if the map values were multiplied by 1.5.  Simulations of the corona and solar wind with the Space Weather Modeling Framework (SWMF) \citep{tothetal2005} have scaled input maps by factors of 2-4 to improve model-observational comparisons \citep{cohenetal2007,jinetal2012,oranetal2015}.

There could be many reasons why various model/map combinations are producing underestimates of $|B_{r IMF}|$, and model parameters can generally be adjusted to open more magnetic flux and increase the value of $|B_{r 1 AU}|$ \citep{stevensetal2012}.  For example, a PFSS model can be made to increase $|B_{r 1 AU}|$ by lowering $R_{SS}$ \citep{leeetal2011}.  However, there is another observational constraint on the models - the predicted open field regions should match CHs observed in emission.  While such comparisons have generally been qualitative, the advent of automated CH detection algorithms \citep{henney_harvey2005,scholl_habbal2008,krista_gallagher2009,lowderetal2014,verbeecketal2014,boucheronetal2016,caplanetal2016} opens the door for more objective comparisons.

\citet{lowderetal2014} performed a comprehensive study of open flux from automatically detected CHs for the 1996-2013 time period, using data from the Solar and Heliospheric Observatory (SOHO) Extreme Ultraviolet Imaging Telescope (EIT), the Solar Terrestrial Relations Observatory (STEREO) Extreme Ultraviolet Imager (EUVI), and the Solar Dynamics Observatory  (SDO) Atmospheric Imaging Assembly (AIA) to detect CHs.  They noted the higher quality of EUVI and AIA images relative to EIT increases the detection of CHs.  They also compared the results to PFSS solutions computed with WSO.  \citet{lowderetal2017} extended this study to investigate the latitude dependence of CHs and contrast the differing behavior of cycle 23 and cycle 24.  We discuss the relationship of our results to \citet[2017]{lowderetal2014} in section~\ref{observation_estimate}.

In this paper, we investigate the open magnetic flux for the time period surrounding July 7-8, 2010 (during Carrington Rotation 2098, June 16--July 13, 2010), employing magnetic maps developed from several instruments and using different map assembly techniques, and computing both PFSS and MHD models.  In section~\ref{maps_models}, we show that the comparison of the predicted open field regions with CHs observed in emission, and the predicted  $|B_{r 1 AU}|$ with in situ spacecraft measurements, together, are powerful constraints on the models and the magnetic maps used to derive the boundary conditions.
We find that no model/map combination can match the inferred $|B_{r IMF}|$ unless the area of their open field regions exceed the CH areas inferred from EUV emission \citep[derived from the automated CH detection scheme described by][]{caplanetal2016}.  In section~\ref{observation_estimate}, we use identified CH boundaries and observatory maps to derive observation-based estimates of the open flux in the corona, and show that these also fall well below the the inferred $|B_{r IMF}|$.  In section~\ref{detection_test}, we employ our CH detection technique on the emission predicted by the MHD model, and show that it captures a large fraction of the open field regions and magnetic flux.  Section~\ref{conclusions} discusses the implications of our results.

\section{Comparison of Maps and Models}
\label{maps_models}

Despite their widespread use for not only scientific, but space weather operational purposes \citep{pizzoetal2011}, magnetic maps from different observatories may agree qualitatively but often disagree quantitatively \citep{rileyetal2014}.   In practice, magnetic maps are made using a variety of methods with differing assumptions.  To better understand how these differences translate into physical solutions, as well as how ``poor'' maps may affect coronal and solar wind model results, a campaign event was organized for the SHINE 2016 workshop.  Prior to the workshop, several global magnetic maps were produced for the same time period using different instruments and different methods, and models were then computed using boundary conditions derived from these maps.  

Full sun maps based on five different observatory magnetogram products were created and supplied for the workshop:  NSO VSM line-of-sight (LOS), NSO Global Oscillation Network Group (GONG) LOS, SOHO Michelson Doppler Imager (MDI) LOS, SDO Helioseismic and Magnetic Imager (HMI) LOS, and SDO HMI vector.  Here we distinguish between two types of magnetic maps created from magnetograms:  Diachronic and Synchronic.
  
Diachronic maps (commonly referred to as synoptic maps) are constructed by projecting full disk magnetograms onto the Carrington (latitude, longitude) frame over the course of a solar rotation.  The construction usually involves the averaging of new magnetograms with earlier data such that each longitude of the map is heavily weighted by the magnetogram(s) taken when that longitude was at disk center.  These maps are the typical product provided by most observatories.

Synchronic maps attempt to approximate the Sun's surface magnetic field at a particular time.  This is obviously difficult, as magnetograms are only observed along the Sun-Earth line at the present time.  Synchronic maps can be constructed simply by inserting a magnetogram from the current date/time into a diachronic map, or, in more advanced approaches, by assimilating magnetograms into a flux transport model that evolves the magnetic field on the unobserved portions of the Sun with known flow and diffusion patterns.  In addition to using diachronic maps from the above observatories, synchronic maps were constructed using the ADAPT model (based on NSO VSM magnetograms) and the LMSAL Evolving Surface-Flux Assimilation Model 
 \citep[ESFAM,][based on SOHO MDI magnetograms]{schrijver_derosa2003}
as well as daily updated synoptic maps for GONG, SDO HMI, SOHO MDI, and NSO VSM (referred to as NSO VSM near real time).  Three different maps from the ADAPT model were used, each a sample  
realization of an ensemble of twelve, differing by the following:  One map included a far-side detection of active region (AR) 11087 observed with GONG helioseismic acoustic holography on July 1 \citep{argeetal2013}, a second map with this same AR included but with the polarity reversed, and a third map included no far-side AR information.    The LMSAL and daily updated maps correspond to 07/08/2010, the ADAPT maps 07/07/2010.  In total, fourteen maps for this time period were used in the results presented in this paper; further details about the maps can be found in the Appendix.
%
%
While the maps were supplied in differing formats and resolutions, we processed all of them as uniformly as feasible (see Appendix) prior to performing the model calculations. 
PFSS calculations with $R_{SS} = 2.0$ and $R_{SS} = 2.5$ were performed for all maps (additional PFSS models were computed for the VSM).  The PFSS  are computed numerically on a nonuniform $151\times 301 \times 602$ ($r,\theta,\phi$) spherical mesh using finite differences and a preconditioned conjugate gradient method \citep{caplanetal2017}.  Thermodynamic MHD models using MAS/CORHEL \citep{lionelloetal2009} were computed on a nonuniform $181\times 251 \times 602$ mesh covering a domain from $1-30R_S$ for selected maps, using the same heating and acceleration parameters for each model.  The MHD results do not depend on the position of the outer boundary, as long as it placed well beyond the MHD fast mode critical point (occurring at about $10-12R_S$ for these simulations).  The solutions were integrated for two days of simulated time until an approximately steady configuration was obtained.  For the MHD results, slight differences were obtained for the open flux (shown in Table 1, where it was computed using Eq.~\ref{eq_phi_open}) depending on the method used for computation.  For example, when the open magnetic flux is computed on the lower boundary (integrating magnetic flux from all field lines that reach the upper boundary), the value obtained is $\sim$1\% less than Table 1; this is due to the presence of a small amount of disconnected flux at some locations in the simulated heliospheric current sheet.  Computing the open flux via  Eq.~(\ref{eq_phi_open}) but using $r = 18.9 R_S$ (above the critical point but below the upper boundary) results in $\sim$3\% larger open flux than Table 1; this is due to the presence of long, closed fields that might eventually reach the upper boundary if the calculation were relaxed even longer.  None of these differences were significant for the results presented here.

\begin{figure}[tb]
 \centering
 \resizebox{0.98\textwidth}{!}{
  \includegraphics{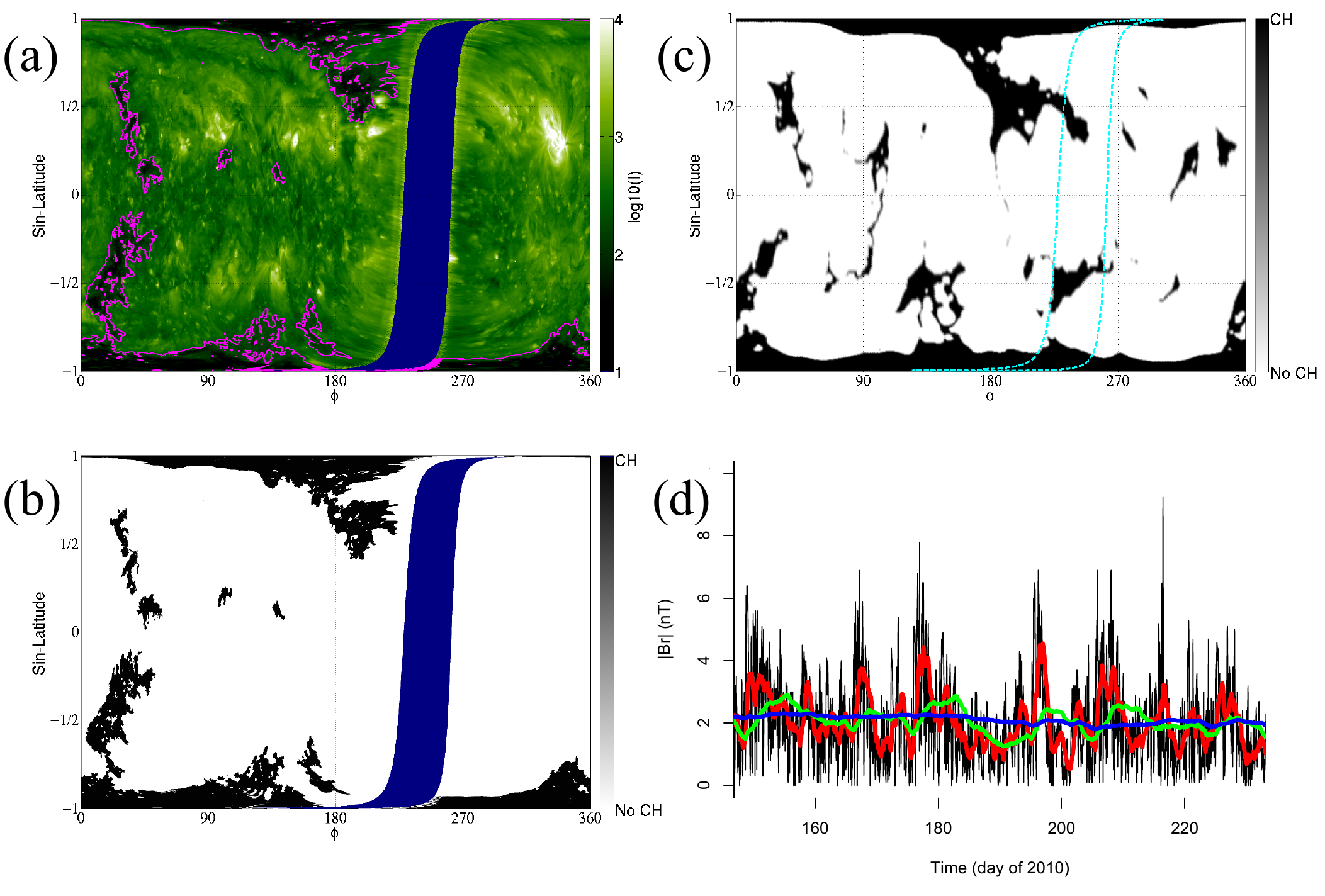}}
\caption{(a) Synchronic EUV map for July 8, 2010 at 18:00, compiled from STEREO A \& B EUVI 195\AA ~and SDO AIA 193\AA ~images.  The magenta lines show the coronal hole detections.  The sector near 270$^\circ$, indicated by the blue swath, was not observed.
(b) The detected coronal holes (black regions) from (a). (c) Open field regions (black) from a thermodynamic MHD model with boundary derived from an ADAPT map. The unobserved region is indicated with cyan dashed lines.  (d) OMNI in situ measurements of $B_r$ for 80 days surrounding the time period of interest.  A 1-hour running average of $|B_r|$ (black line), 1-day running average (red),  7-day (green) and Carrington rotation average (blue) are shown.
	\label{fig_ch_interplanetary} }
 \end{figure}

As discussed in the introduction, all models can be compared with two basic observations:  (1) open field areas, as deduced from CH detection, and (2) the approximate open magnetic flux in the heliosphere, as estimated from in situ spacecraft measurements.  For (1), we take advantage of a recently-developed database of synchronic EUV and automatically detected CH boundaries \citep{caplanetal2016} that is publicly available ({\tt www.predsci.com/chd}).  This provides a digital representation of the CHs that can be compared quantitatively with models.  The data sources are the STEREO EUVI 195\AA~and SDO AIA 193\AA ~images.  \citet{caplanetal2016} describe the techniques used to construct synchronic EUV maps from these images that approximate the view from disk center and a single instrument for all locations, as well as the CH detection method.  Figure~\ref{fig_ch_interplanetary}(a) shows a synchronic EUV map for 7/8/2010 and Figure~\ref{fig_ch_interplanetary}(b) shows the corresponding CH detection.  The total CH area detected was $7.5\times10^{21}$cm$^2$ (this excludes the region that was not observed by the STEREO or SDO spacecraft, seen as a blue swath in Figures~\ref{fig_ch_interplanetary}(a-b)).  We refer to this as the ``observed'' area, but it is important to remember that all CH detection algorithms have adjustable parameters, and the area may depend on these.  Figure~\ref{fig_ch_interplanetary}(c) shows the open field regions from an example map/model combination (the MHD model using the ADAPT map with the farside AR included), calculated by tracing field lines from the solar surface and marking cells as open (black) if they reach the upper boundary and closed (white) if they return to the solar surface.  Comparing Figure~\ref{fig_ch_interplanetary}(c) with 
Figures~\ref{fig_ch_interplanetary} 1(a)-(b), we see that the model produces a larger CH area than was detected in EUV.  The results for all of map/model combinations are described below.

\begin{table}[tbp] 
\centering
\begin{tabular}{ |p{3.7cm}||p{1.9cm}|p{2.2cm}|p{2.5cm}|p{2.2cm}|p{2.7cm}|  }
 \hline
~~~~~~~~~~~1 & ~~~~~~~~2 & ~~~~~~~3 & ~~~~~~~~~4 & ~~~~~~~5 & ~~~~~~~6 \\
 \hline
Magnetic Map   & Unsigned Flux ~~~~~($10^{22}$ Mx)&Average Polar Field (G) South/North & Model  &Open~Field Area  ~~~~~~~(difference) ($10^{21}$cm$^2$)&Open Flux ($B_r$ at 1 AU, nT) \\
\hline
 Observed  & &  &  & 7.6 (EUV) & 1.7-2.2~(OMNI)\\
 \hline
ADAPT, Far Side&17.9&   3.1 (S)&PFSS,~2.5$R_{SS}$   & 5.8~(-1.8) & 0.75  \\
 (NSO VSM  & &   -2.6 (N)&PFSS,~2.0$R_{SS}$  & 6.9~(-0.7)& 0.94 \\ 
magnetograms) & &    &MHD& 8.9~(+1.3)& 1.35\\ 
\hline
 ADAPT,~Far~Side,  &17.6&   3.1 (S) &PFSS,~2.5$R_{SS}$&  6.3~(-1.3) & 0.82\\
 AR~polarity~reversed & &   -2.6 (N)&PFSS,~2.0$R_{SS}$   & 7.4~(-0.2)& 1.03\\ 
 & &      &MHD & 8.7~(+1.1)& 1.33\\ 
  \hline
 ADAPT,~No~Far~Side&14.8&   3.1 (S) &PFSS,~2.5$R_{SS}$ & 6.1~(-1.5) & 0.76\\
   & &   -2.6 (N) &PFSS,~2.0$R_{SS}$& 7.1~(-0.5)& 0.94\\ 
   & &    &MHD& 9.3~(+1.7)& 1.28\\ 
   \hline 
GONG Daily&11.4&   2.6 (S) &PFSS,~2.5$R_{SS}$ & 6.0~(-1.6)& 0.62\\
  Synoptic  & &   -2.4 (N) &PFSS,~2.0$R_{SS}$& 7.0~(-0.6) & 0.75\\ 
   \hline 
GONG Synoptic  &11.3&   2.6 (S) &PFSS,~2.5$R_{SS}$ & 6.3~(-1.3)& 0.64\\
   & &   -2.4 (N) &PFSS,~2.0$R_{SS}$& 7.3~(-0.3) & 0.77\\ 
   \hline 
   HMI LOS &12.9&   2.8 (S) &PFSS,~2.5$R_{SS}$ & 5.8~(-1.8)& 0.66\\
   Daily Updated & &   -2.7 (N) &PFSS,~2.0$R_{SS}$& 6.7~(-0.9)& 0.79\\ 
    \hline 
   HMI LOS Synoptic &13.9&   2.9 (S) &PFSS,~2.5$R_{SS}$ & 5.4~(-2.2)& 0.65\\
   & &   -2.7 (N) &PFSS,~2.0$R_{SS}$& 6.3~(-1.3)& 0.79\\ 
   \hline
    HMI~Vector~Synoptic &15.1&   3.5 (S) &PFSS,~2.5$R_{SS}$ & 5.4~(-2.2)& 0.80\\
   & &   -3.7 (N) &PFSS,~2.0$R_{SS}$& 6.3~(-1.3)& 0.96\\  
   \hline
    LMSAL~ESFAM &13.2&   3.9 (S) &PFSS,~2.5$R_{SS}$& 4.3~(-3.3)& 0.64\\
  (MDI magnetograms) & &   -2.4 (N) &PFSS,~2.0$R_{SS}$& 5.3~(-2.3)& 0.78\\ 
  & &    &MHD & 7.8~(+0.2)& 1.12\\ 
   \hline 
    MDI Daily updated &18.4&   3.5 (S) &PFSS,~2.5$R_{SS}$ & 4.8~(-2.8)& 0.75\\
   & &   -3.2 (N) &PFSS,~2.0$R_{SS}$& 5.7~(-1.9)& 0.92\\ 
   \hline
    MDI Synoptic &18.2 &   3.3 (S)&PFSS,~2.5$R_{SS}$& 5.1~(-2.5)& 0.73\\
   & &   -3.2 (N) &PFSS,~2.0$R_{SS}$ & 5.9~(-1.7)& 0.90\\ 
   \hline 
  VSM Synoptic &16.3&   3.4 (S) &PFSS,~2.5$R_{SS}$& 5.5~(-2.1)& 0.79 \\
   & &   -3.3 (N) &PFSS,~2.0$R_{SS}$& 6.4~(-1.2)& 0.96\\
   & &   &PFSS,~1.4$R_{SS}$& 10.7~(+3.1)& 1.60\\
    & &   &PFSS,~1.3$R_{SS}$& 12.8~(+5.2)& 1.91\\
 \hline 
  VSM Synoptic  &17.8&   3.3 (S) &PFSS,~2.5$R_{SS}$ & 5.3~(-2.3)& 0.83\\
  (extrapolated polar & &   -3.7 (N) &PFSS,~2.0$R_{SS}$& 6.2~(-1.4)& 1.01\\ 
   fields) & &    &MHD& 7.9~(+0.3)& 1.38\\ 
   \hline 
 VSM~Near~Real~Time  &16.3 &   3.1 (N) &PFSS,~2.5$R_{SS}$ & 5.4~(-2.2)& 0.77\\
   & &  -3.5 (S) &PFSS,~2.0$R_{SS}$ & 6.4~(-1.2)& 0.95\\ 
   \hline 
\end{tabular}
\caption{Summary of results from all of the model/map combinations. \label{table_map_model}}
\end{table}

To estimate the average interplanetary $|B_{r IMF}|$ during this time period, we obtained one-hour averaged OMNI in situ measurements of $B_r$ and computed the absolute value.  Figure~\ref{fig_ch_interplanetary}(d) shows   the 1-hour data (black), a 1-day running average (red), a 7-day average (green), and a Carrington rotation average (blue), for an 80 day period (5/30-8/18/2010).  The average value during the plotted interval is 2.19 nT (nano Tesla); the average for CR2098 alone is 2.07 nT.  However, these values could be an overestimate of the interplanetary magnetic flux.  \citet{lockwoodetal2009} argued that kinematic effects can create longitudinal structures in the solar wind where the IMF folds back on itself \citep{crookeretal2004}, and this can lead to an "over-counting" of magnetic flux from $|B_{r IMF}|$ measurements \citep{owensetal2013}.  These inverted magnetic structures show the signature of an HCS crossing ($B_r$ reverses sign) but suprathermal electrons travel radially inward along the field (typically, these electrons travel outward along open field lines).  To account for this effect, we examined the 27 day time period for CR2098 in ACE measurements and found 88 hours of inverted magnetic flux.  Removing this flux drops the average of $|B_{r IMF}|$ for this time period from 2.07nT to 1.69nT.  We also obtained the daily averaged $B_r$ from OMNI for the same time period as shown in 
Figure~\ref{fig_ch_interplanetary}(d) and found the average of $|B_{r IMF}|$ to be 1.67nT.  This latter estimate is likely to be low, because at a daily time-averaging interval, $B_r$ measured near the HCS will tend to cancel, reducing the value.  Using these three different estimation methods, we conclude that the average interplanetary magnetic flux for this time period corresponds to a value of $|B_{r IMF}|$ between 1.7 and 2.2nT.

Table~\ref{table_map_model} summarizes the results for all of the map/model combinations, and their comparison with observations.  The first column of the table identifies the map, column 2 lists the unsigned magnetic flux for the entire map (= $R_S^2 \int_{0}^{4\pi} |B_r(R_S,\theta,\phi)|d\Omega$), and column 3 shows the integrated magnetic flux 
within 25$^\circ$ of the pole expressed as an average radial field strength (= $(R_S^2/A_n) \int_{0}^{2\pi} \int_{0}^{5\pi/36}B_r(R_S,\theta,\phi)\sin\theta d\theta d\phi$, north; = $(R_S^2/A_s) \int_{0}^{2\pi} \int_{31\pi/36}^{\pi}B_r(R_S,\theta,\phi)\sin\theta d\theta d\phi$, south; $A_n = A_s = R_S^2 \int_{0}^{2\pi} \int_{0}^{5\pi/36}\sin\theta d\theta d\phi$  $= 0.187\pi R_S^2$).  Column 4 identifies the model calculations performed for each map.  Column 5 shows the integrated open field area (disregarding the region not observed by SDO and the STEREOs) for each map/model; in parenthesis is the difference between this area and the CH area deduced from observed emission.  This simple metric generally underestimates the discrepancies between two CH maps, because disagreements (to much open area in one region, too little in another) can cancel.  However, it is sufficient for our purposes here, as we are primarily interested in constraining how much open flux can be produced by a map/model while still remaining consistent with emission observations.  Column 6 shows the equivalent open flux ($|B_{r 1 AU}|$ , computed from eq.~\ref{eq_phi_open}) for each map/model.  While all of the maps/models approximate the global solar magnetic field for the same time period, considerable variability in the results is seen.

The striking result from Table~\ref{table_map_model} is that for all MHD map/models and all PFSS map/models with $R_{SS}=2.0$ \& $2.5$, $|B_{r 1 AU}|$ falls well below the observed range of $|B_{r IMF}|$ (1.7-2.2nT).  The $R_{SS}= 2.5$ PFSS models clearly underestimate both the open field area and $|B_{r IMF}|$, implying that the magnetic field is opening much lower in the corona during this time period.  The open field areas of the $R_{SS}= 2.0$ PFSS models are generally much closer to (but smaller than) the observed CH area, but their values for $|B_{r 1 AU}|$ are still much smaller than $|B_{r IMF}|$.  The MHD model $|B_{r 1 AU}|$ values come closest to $|B_{r IMF}|$, but the open field areas all exceed the observed CH area.  The greater opening of flux is not an inherent property of MHD, but rather is related to the model parameters, such as the heating model.  While the relationship between field opening and model parameters is more complex in MHD than in PFSS (the field does not open at one height, and the length scale of heating deposition is as important as the magnitude in determining coronal structure), increasing the open flux produced by the model generally requires increasing the open field area, just as with the PFSS.  The introduction of shear and/or twist in a model can also cause more flux to open  \citep[e.g.,][]{rileyetal2006,edwardsetal2015}, but this will also increase the open field area predicted by the model.

\begin{figure}[tb]
 \centering
 \resizebox{0.98\textwidth}{!}{
 \includegraphics{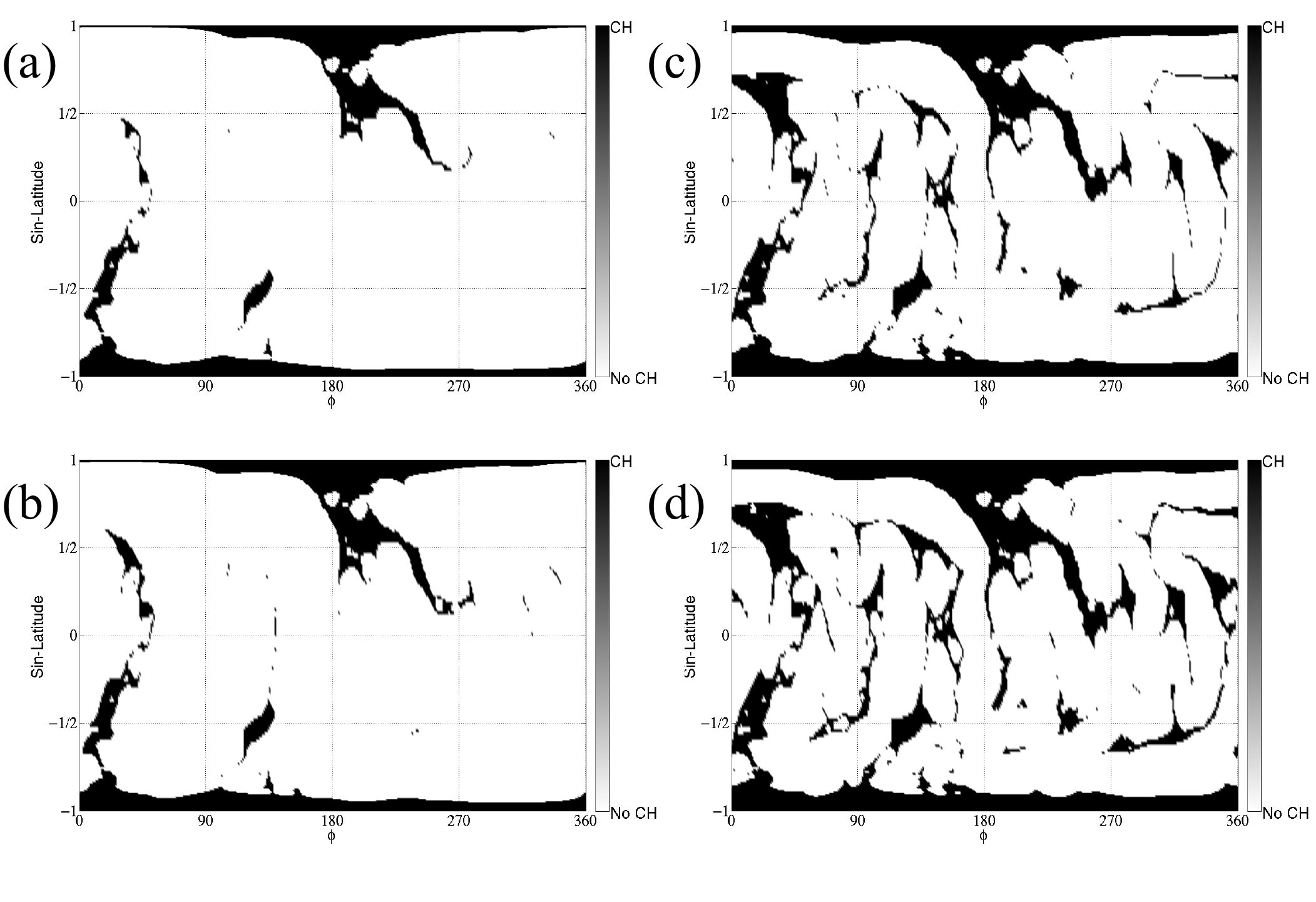}}
 \caption{
Open field regions (black) for four PFSS models using the NSO VSM CR2098 map for the boundary condition.  (a) $R_{SS} = 2.5R_S$.  (b) $R_{SS} = 2.0R_S$.  (c) $R_{SS} = 1.4R_S$.  (d) $R_{SS} = 1.3R_S$.
 } \label{fig_ch_4SS_models}
 \end{figure}

We can ask the question, for a given map, how large an area has to be opened in order to match $|B_{r 1 AU}|$?  The answer is shown in the table entry for VSM synoptic.  
Using the PFSS model and lowering $R_{SS}$ to 1.4$R_S$, the model yields $|B_{r 1 AU}|$ =1.60nT, which still falls outside the observed range for $|B_{r 1 AU}|$, and the open field area is now 41\% greater than observed.   Further lowering $R_{SS}$ to 1.3$R_S$, we obtain $|B_{r 1 AU}| = 1.91$nT.  This value now falls in the observed range but the open field area is 68\% greater than observed.  Figure~\ref{fig_ch_4SS_models} shows the open field regions for the four PFSS models computed using the VSM synoptic map ($R_{SS} = 2.5, 2.0, 1.4, 1.3R_S$)  The $R_{SS}  = 2.0R_S$ case (Figure~\ref{fig_ch_4SS_models}(b)) appears visually closest to the CHs in the synchronic (Figure~\ref{fig_ch_interplanetary}(a-b) and diachronic (Figure~\ref{fig_ch_magnetogram_flux}(a-b)) EUV map/detections.  The $R_{SS}  = 1.4$ and $1.3R_S$ models (Figure~\ref{fig_ch_4SS_models}(c-d)) are visually inconsistent with the EUV maps and detections.  No map/model combination is consistent with our two constraints.

\section{An Observation-derived Estimate of Open Flux}
\label{observation_estimate}
Our digital CH database allows us to estimate the open magnetic flux directly from solar observations.  We can calculate the open flux for a given magnetic map by overlaying the CH map over the magnetic map and integrating the magnetic flux in each hole individually.  To obtain an estimate of the average open magnetic flux over a solar rotation, the time sequences of the CH observations should approximately correspond to the timing of the magnetic observations.   Therefore, the most straightforward approach is to use diachronic maps provided by observatories for the magnetic data, and to develop a diachronic CH map coincident with the magnetic maps.  We developed such a map from our EUV/CH database by weighting each synchronic map with a longitudinal Gaussian centered on the (Earth based) Carrington longitude at the time of the map.  We chose a Gaussian full-width-half-max value of one degree, and the weighted maps were combined to create the final synoptic map.  This was done independently for the EUV and CH maps. Because of the weighting, the weighted EUV/CH map data originates mostly from the AIA instrument but, due to the nature of the merged synchronic maps, there is some contribution from STEREO A/B data near the polar regions that are unobserved by AIA.

\begin{figure}[tb]
 \centering
 \resizebox{0.98\textwidth}{!}{
 \includegraphics{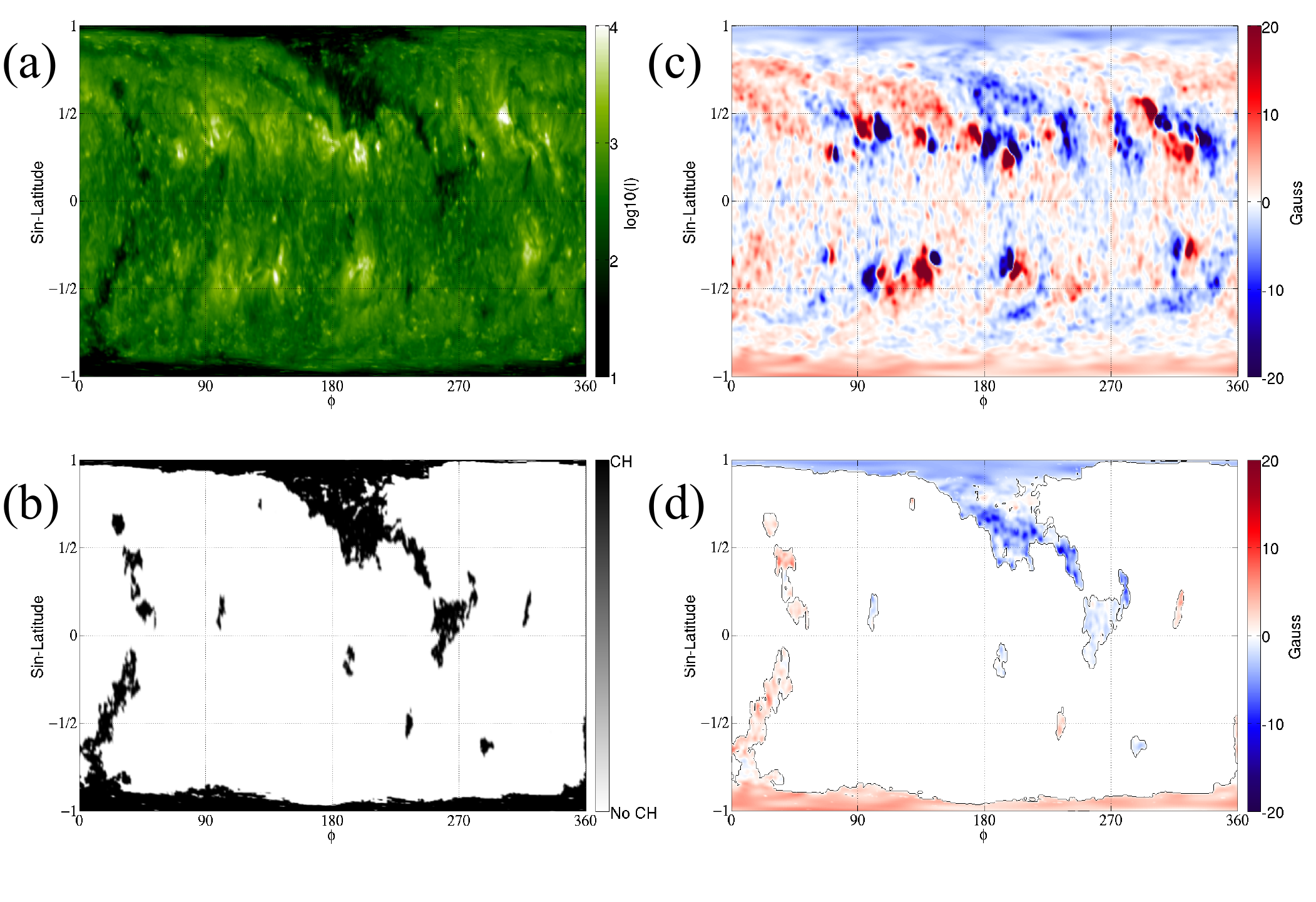}}
 \caption{
(a) Diachronic EUV map for CR2098 (6/16-7/13/2010), constructed predominantly from AIA 193\AA~ images, plotted as sine(latitude) vs.\ longitude  (b) Corresponding CH detections for (a), plotted in the same format.  (c) $B_r$ at the photosphere derived from the LOS field, for data from the NSO VSM for this Carrington rotation.  (d) The magnetic field in the CH, obtained by overlaying (b) with (c).
 } \label{fig_ch_magnetogram_flux}
 \end{figure}

The resulting EUV map is shown in Figure~\ref{fig_ch_magnetogram_flux}(a), and the CH detections from this map are shown in Figure~\ref{fig_ch_magnetogram_flux}(b).  Figure~\ref{fig_ch_magnetogram_flux}(c) shows the NSO VSM diachronic map for $B_r$ at the photosphere.  The overlaying of (b) on (c) yields Figure~\ref{fig_ch_magnetogram_flux}(d), a map of the magnetic field in each CH for this time period.

Using the method shown in Figure~\ref{fig_ch_magnetogram_flux}, we calculated the open flux using detected CHs in Figure~\ref{fig_ch_magnetogram_flux}(b) and the five diachronic observatory maps described in section~\ref{maps_models}.  
The results are summarized in Table~\ref{table_observed_open}.  The second column of the table shows the unsigned magnetic flux in all of the CHs for the different maps.  The third column of Table~\ref{table_observed_open} is computed by integrating the signed flux in each CH individually, then adding the absolute value of each of these together.  The percentage of this flux compared to the unsigned flux (column 2) is shown in paranthesis.  Our CH detection technique relies only on EUV measurements and does not use the magnetic flux; the close match between the signed and unsigned flux indicates that the technique is indeed identifying predominantly unipolar regions as CHs.  A check on the technique is shown in column 4 of Table~\ref{table_observed_open}.  Unlike the case for the models, the total flux from the positive and negative detected CHs is not guaranteed to balance, if, for example, some CHs are missed by the technique.  The relatively low flux balance error (sum of all the signed fluxes divided by the unsigned flux) shows that the detected CHs have nearly equal amounts of positive and negative flux.  Column 5 of the table shows the predicted $|B_{r IMF}|$ computed from column 3.  For all of the maps, the open flux predicted to arise from the CHs is well below that inferred from in situ measurements.
\begin{table}[h] 
\centering
\begin{tabular}{ |p{3.9cm}||p{3cm}|p{3.5cm}|p{2.5cm}|p{2cm}| }
 \hline
~~~~~~~~~~~1 & ~~~~~~~~~~~2 & ~~~~~~~~~~~3 & ~~~~~~~~4  & ~~~~~5 \\
 \hline
Observatory Map   & Total Unsigned Flux in all CHs ($10^{22}$ Mx)& Total~Open~Flux in~all~CHs~($10^{22}$ Mx) ~~~~~~~~~~~~~~~(\%~of~unsigned~flux) &Flux Balance Error &Open Flux ($B_r$ at 1 AU, nT) \\
 \hline
GONG synoptic &1.91 &1.89 (99.0\%) & 3.6\%  & 0.67 \\
\hline
HMI LOS synoptic&2.09 &2.01 (96.2\%) & 2.9\%  & 0.71 \\
\hline
HMI Vector synoptic&2.56 &2.47 (96.5\%) & 5.8\%  & 0.88 \\
\hline
MDI synoptic &2.49 &2.38 (95.6\%) & 5.6\%   & 0.85\\
\hline
VSM synoptic &2.55 &2.51 (98.4\%) & 5.7\%  & 0.89 \\
\hline
\end{tabular}
\caption{Open fluxes computed using observed coronal holes and five diachronic maps.  \label{table_observed_open}}
\end{table}

\citet[2017]{lowderetal2014} used automated CH detection to compute magnetic flux in coronal holes over many years, including this time period.   There are significant differences between our method and theirs; \citet{lowderetal2014}'s CH detection uses a single threshold on partitioned subarrays of EUV images, and incorporates magnetic field observations (relative unipolarity of regions).  They compute the unsigned flux in CHs over the whole Sun.  We perform substantially more processing of the EUV images (e.g., correcting for limb-brightening) prior to CH detection, allowing the use of a global two-threshold method \citep{caplanetal2016}.  We compute the signed fluxes in each CH individually, which is much less sensitive to map resolution than the unsigned flux.  \citet{lowderetal2014}'s detected CHs appear to be larger at low latitude for CR2098 (Figure 10 of their paper) than in our method; comparison of the EUV map with the CH for this rotation indicates that at least some of the areas they identify as CH are more consistent with quiet Sun.  Nevertheless, their values for the open flux during this time period ($\sim$2$\times10^{22}$ Mx) are approximately consistent with ours.  \citet{lowderetal2017} also estimated the interplanetary open flux from OMNI measurements.  While they do not describe the averaging technique used to obtain the interplanetary open flux, the values they obtain are greater by a factor of 2 or more than the open flux they estimate from CHs for much of cycle 23 and all of cycle 24, including the time period studied here.  Their result suggests that the underestimate in open magnetic flux we have identified for this time period is a pervasive issue.

\section{Coronal Hole Detection Applied to Simulated Emission}
\label{detection_test}

Given that the regions identified as CHs in emission apparently cannot account for the measured interplanetary flux for this time period, it is important to assess how well the CH detection method identifies CHs and their corresponding open flux.  The thermodynamic MHD model allows us to simulate emission as it would be observed in different instruments \citep{lionelloetal2009}.  As the  true open field regions and open magnetic flux for the model are known, we tested the detection technique, by applying it to simulated emission from the MHD model that used the NSO VSM CR2098 map as a boundary condition.  

To perform the test, we computed the simulated emission using radial lines of sight at each pixel, as this roughly matches the way the diachronic CH map of 
Figure~\ref{fig_ch_magnetogram_flux}(a) (heavily weighted by disk center observations) was calculated; however, the use of radial lines of sight is likely to make the detection of CHs at high latitudes more accurate then in the case of real STEREO/SDO observations.  The parameters for the CH detection used on the synchronic EUV maps were optimized for the emission levels in the STEREO/SDO data in the 2010-2014 time period.  The emission in the MHD model differs quantitatively from the actual Sun, so for our detection test, we optimized the detection parameters for the simulated emission.

\begin{figure}[tb]
 \centering
 \resizebox{0.98\textwidth}{!}{
 \includegraphics{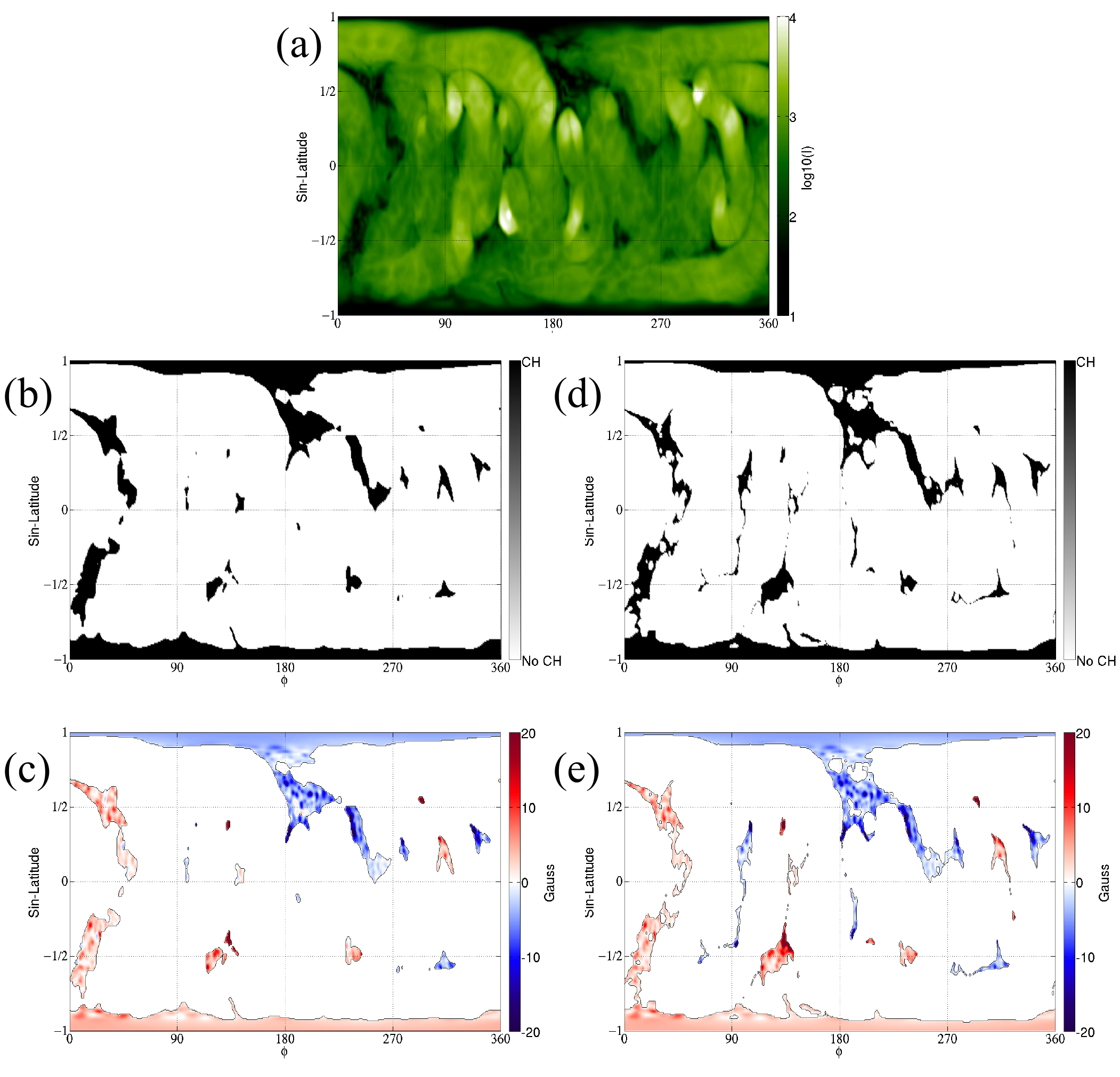}}
 \caption{
(a) Simulated AIA 193\AA ~map from the thermodynamic MHD model using the NSO VSM CR2098 map for the boundary condition.  This map is used to test the CH detection method. (b) The CHs (black regions) identified when the CH detection method is applied to the simulated data.  (c) The magnetic field in the CH, obtained by overlaying (b) with the NSO VSM map (as was done in Figure~\ref{fig_ch_magnetogram_flux}).  (d) The true open field regions for the model.  (e) The true open magnetic field in the model. 
 } \label{fig_ch_detection_model}
 \end{figure}

We note that visually the dark emission regions in the simulation 
(Figure~\ref{fig_ch_detection_model}a) are similar to the observed (Figure~\ref{fig_ch_magnetogram_flux}a) but generally larger.  Figure~\ref{fig_ch_detection_model}(b) shows the CH regions from the simulation identified by the CH detection method, and Figure~\ref{fig_ch_detection_model}(c) shows the magnetic flux identified as open.  The CH area detected in 
Figure~\ref{fig_ch_detection_model}(b) is $8.53\times10^{21}$cm$^2$ and the absolute value of the signed flux in Figure~\ref{fig_ch_detection_model}(c) is $3.42\times10^{22}$ Mx.  Comparing these values with the true open field region area from the model ($8.82\times10^{21}$cm$^2$, shown in Figure~\ref{fig_ch_detection_model}(d)) and the true open flux ($3.87\times10^{22}$ Mx,
 shown in Figure~\ref{fig_ch_detection_model}(e)), we find that the detection technique accounts for 96.8\% of the CH area and 88.4\% of the flux.  At least for the case of the model, under-detection of CHs in emission results in missing a relatively small fraction of the open flux.  Further tests of our detection technique using simulated emission are planned for the future.

It is interesting to note that while the CH detection method underestimated CH area by only 3.2\%, this translated into a much larger open flux error (11.6\%).  The reason for this can be seen by comparing Figure~\ref{fig_ch_detection_model}(c) and (e).  Some localized, strong concentrations of magnetic flux in mid-latitude regions are either missing or not fully captured by the CH detection.  These flux concentrations originate in open fields emanating from the edges of active regions.  This could perhaps give clues to where the missing open flux resides, as we discuss in the following conclusions section.  

\section{Conclusions}
\label{conclusions}
We surveyed fourteen magnetic maps created from five different magnetogram products representing the time period surrounding early July, 2010, computing solutions with PFSS and MHD models.  As this time period occurred early in cycle 24, the Sun's magnetic flux was evolving relatively slowly and the coronal configuration was near minimum.  Therefore, we would expect the standard paradigm of coronal structure to hold and that the open magnetic flux would primarily arise from polar coronal holes and their equatorial extensions.  However, we found that all of the model/map combinations underestimate the interplanetary magnetic flux for this time period (inferred under the assumption that this can be estimated from OMNI in situ measurements), unless the open field regions of the model exceed the CH area that is inferred from EUV emission.  When we used the detected CH areas together with observatory maps (bypassing the requirement of a model), all cases underestimated the interplanetary flux by close to a factor of two or more.

There are two broad categories of resolutions for this underestimate of the open flux:   (1) Either the observatory maps are underestimating the magnetic flux, or (2) a significant portion of the open magnetic flux is not rooted in regions that are dark in emission.  While (1) could be occurring on large portions of the solar surface for all of the observatories, it seems unlikely.  The maps employed in this study incorporate magnetograms that were derived from instruments measuring different formation lines, and detecting the field at different depths in the photosphere; why all of these different instruments would underestimate the field is unclear.  On the other hand, the poles of the Sun are poorly observed, and it is possible that the polar magnetic flux could be significantly underestimated near solar minimum, as implied by Hinode Solar Optical Telescope observations \citep{tsunetaetal2008}.  \citet{tsunetaetal2008} estimated that the concentrated kilogauss patches that they observed would give the equivalent flux of a 10 G field over the region 20$^\circ$ from the pole; integrating the polar field from the NSO VSM synoptic map for that time period (3/16/2007, CR2054) yields $\sim$6 G.
A similar underestimate in the polar fields for the time period studied here might account for the missing flux.  Future Solar Orbiter observations in the later part of the mission (when the spacecraft is well out of the ecliptic plane) could provide a more definitive view of the contribution of polar fields to the interplanetary magnetic flux.

At this point, possibilities in category (2) are speculative, and at least to some degree depart from the standard paradigm.  They could range from issues with CH detection to invocations of time-dependent effects.  An example of the first possibility would be active regions contributing more open flux, but where the footpoints are obscured by bright emission  (requiring a vastly greater contribution than was found in our model test shown in Figure~\ref{fig_ch_detection_model}).   The latter case could be related to the possibly dynamic origin of the slow solar wind, a subject of considerable controversy \citep[e.g.,][]{abboetal2016}.  A particular example is the S-web model \citep{antiochosetal2011,linkeretal2011,titovetal2011}, which argues that an important portion of the slow wind arises from interchange reconnection between closed and open fields \citep{fisketal98}.  If this is the case, then perhaps the regions bounding coronal holes contains significant amounts of open field intermixed with closed field and are not dark in emission.  Demonstration of this effect in emission by a model/simulation would be a first step towards investigating the viability of this idea.  

We note that the excess flux produced by CMEs cannot by itself resolve the open flux problem.  CMEs are believed to contain magnetic flux connected back to the Sun at both ends \citep[as evidenced by counterstreaming electrons,][]{goslingetal1987}.  These long field lines behave just like open field lines in terms of EUV emission close to the Sun and their footpoints would presumably be dark; some other reason would need to be invoked as to why they could be embedded in bright regions.  If there were large amounts of disconnected flux in the heliosphere, it could account for the missing flux.  While this doesn't seem to be directly ruled out by present observations, it is generally considered unlikely \citep{crooker_pagel2008}.

In closing, we note that we have focused on the issues of accounting for the open flux for a single, well-observed time period.  We believe these discrepancies are likely to be ubiquitous, and so should be investigated further, perhaps with more coordinated observing time periods.  The comparison of different CH detection techniques for periods of interest (particularly during the 2010-2014 time period, with EUV coverage of nearly the entire Sun from the STEREO and SDO spacecraft) could be especially useful.
\acknowledgements
This work was supported by AFOSR contract FA9550-15-C-0001, NASA HSR grant NNX17AI29G, NASA HGI grant NNX17AB78G, the LWS TR\&T program, the STEREO SECCHI contract to NRL (under a subcontract to PSI), and NSF's FESD program.  Computational resources were provided by NSF's XSEDE and NASA's NAS.  We thank Xudong Sun of Stanford for providing polar filling for the HMI and MDI maps.  We also wish to acknowledge the SHINE 2016 workshop, which was the genesis of this collaboration.

\appendix
\section{Map Processing}
The maps provided for the SHINE 2016 workshop specified $B_r$ at the photosphere with differing resolution and format, but we adopted a standard pipeline for processing and preparing them for model input.  The NSO GONG and NSO SOLIS VSM CR2098 synoptic maps were supplied as standard products from the NSO web site (http://gong.nso.edu/ and http://solis.nso.edu/index.html), as was the NSO VSM near-real-time map and the GONG daily map; all were provided at 180$\times$360 resolution (sine-latitude vs. longitude).  The SOHO MDI CR2098 synoptic map was obtained from the Stanford website (http://sun.stanford.edu/synop/) at 1080$\times$3600 sine-latitude vs. longitude.  The SDO HMI LOS and vector synoptic maps were supplied by Stanford at 1440$\times$3600 sine-latitude vs. longitude, similar to the products available at the Joint Science Operations Center (JSOC; http://jsoc.stanford.edu/), except the polar fields were filled by the technique described by \citet{sunetal2011}, updated for HMI maps (Sun et al., in preparation).  The creation of full-sun HMI vector maps, including the disambiguation of the transverse field, is described by \citet{liuetal2017}.  The MDI daily map was also supplied at 1440$\times$3600 sine-latitude vs. longitude resolution, while the HMI daily map was supplied at 180$\times$360 latitude vs. longitude; both used the same polar filling technique as the synoptic HMI maps.  The ADAPT and ESFAM maps were provided at 180$\times$360 resolution latitude vs longitude resolution.  We note that for the ADAPT and ESFAM models, the polar fields are not assimilated from magnetograms but arise from the flux transport calculation over many rotations.  For all of the LOS maps, $B_r$ was calculated from the LOS field under the frequently applied assumption that the field is predominantly radial at the depth that it is measured in the photosphere \citep{wang_sheeley1992}.  For the HMI vector map, $B_r$ was provided directly.

Each map was re-interpolated to a uniform, $300\times600$ grid in latitude/Carrington longitude using an integral (flux) preserving interpolation scheme, including the supplied polar field values.  The  VSM Near Real Time map had missing polar values, and these were corrected using our pole fitting/filling algorithm, which replaces data within 23 degrees of each pole using an extrapolation based on data between 23 and 40 degrees of the pole \citep{linkeretal2013}. In addition, we also developed an alternative VSM CR2098 synoptic map (VSM synoptic extrapolated polar fields in Table 1) using pole fitting.  After interpolation, all maps were flux-balanced to enforce $\nabla\cdot \mathbf{B}=0$ using a multiplicative factor and then smoothed to match the final grid resolution. The smoothing was done by advancing a standard diffusion operator over the sphere such that the final diffused length scale was approximately the equatorial cell size. To safely ensure smoothness near the pole (within 30 degrees), the diffusion coefficient slowly increases by a factor of two at the pole.

All of the maps used in this paper, as well as the results shown in all of the figures, will be available upon publication at http://www.predsci.com/open\_flux\_problem.

\bibliographystyle{apj}
\bibliography{open_flux}

\newpage

\end{document}